\documentclass[10pt,twocolumn]{IEEEtran}
\usepackage{amsmath}
\usepackage{amsthm}
\usepackage{amsfonts}
\usepackage{amssymb}
\usepackage{mathrsfs}
\usepackage{cases}
\usepackage{cite}
\usepackage{graphicx}
\usepackage{mathrsfs}
\usepackage{subfigure}
\usepackage{epstopdf}
\usepackage{color}
\usepackage{bm}
\usepackage{caption}
\allowdisplaybreaks[4]
\theoremstyle{remark}

\begin{document}

\title{\LARGE Sequential Anomaly Detection Against Demodulation Reference Signal Spoofing in 5G NR 
\thanks{
S.-D. Wang and H.-M. Wang are with the School of Information and Communication Engineering, Xi’an Jiaotong University, Xi’an, 710049, Shaanxi, China (e-mail: xjtuwsd@stu.xjtu.edu.cn; xjbswhm@gmail.com).
}
\thanks{
C. Feng and V. C. M. Leung  are with the Department of Electrical and Computer Engineering, University of British Columbia, Vancouver, BC V6T 1Z4, Canada (e-mail: chen.feng@ubc.ca; vleung@ieee.org).
}
\author{ Shao-Di Wang, Hui-Ming Wang,~\IEEEmembership{Senior Member,~IEEE}, \\Chen Feng,~\IEEEmembership{Member,~IEEE}, and Victor C. M. Leung,~\IEEEmembership{Fellow,~IEEE}}}

\maketitle

\begin{abstract} 
In fifth generation (5G) new radio (NR), the demodulation reference signal (DMRS) is employed for channel estimation as part of coherent demodulation of the physical uplink shared channel. However, DMRS spoofing poses a serious threat to 5G NR since inaccurate channel estimation will severely degrade the decoding performance. In this correspondence, we propose to exploit the spatial sparsity structure of the channel to detect the DMRS spoofing, which is motivated by the fact that the spatial sparsity structure of the channel will be significantly impacted if the DMRS spoofing happens. We first extract the spatial sparsity structure of the channel by solving a sparse feature retrieval problem, then propose a sequential sparsity structure anomaly detection method to detect DMRS spoofing. In simulation experiments, we exploit clustered delay line based channel model from 3GPP standards for verifications. Numerical results show that our method outperforms both the subspace dimension based and energy detector based methods.
\end{abstract}

\section{Introduction}
\label{Sec:Introduction}
Fifth generation (5G) new radio (NR) is a new wireless access technology which is developed by the 3rd Generation Partnership Project (3GPP) to meet the diverse performance requirements by various use cases. 5G NR physical layer consists of several physical channels and signals, most of which are vital to the operation of the air interface [1]. Among them, the demodulation reference signal (DMRS) associated with the physical uplink shared channel (PUSCH) serves an important role in data transmission from users to enable coherent demodulation. 

However, an intelligent adversary can be protocol-specific to launch more effective attacks. For instance, an intelligent adversary can attack the PUSCH by sending the same DMRS as the legitimate user, referred to as DMRS spoofing. Different from random jamming, spoofing is more intelligent and destructive [2], [3]. The basic idea of the spoofing is that the attacker tries to masquerade as the legitimate transmitter by sending a fake information to the receiver. Once 5G NR network suffers DMRS spoofing, the channel estimation errors for coherent detection of data symbols will severely degrade the decoding performance. Thus, an effective and reliable DMRS spoofing detection is urgently required for 5G NR.

Recently, many detection methods have been proposed for pilot spoofing, nevertheless, existing methods are not suitable for DMRS spoofing detection. For instance, 1) \emph{communication protocol (CP) based method} [4], [5] is mainly based on the private communication protocol by introducing extra randomness, which is impractical. This is because the technical specification of DMRS in 5G NR has been standardized by 3GPP; 2) \emph{auxiliary transmission (AT) aided method} [6], [7] has a high startup cost (training time), because it requires the base station (BS) to make the decision by collecting all the observations due to the introduction of auxiliary transmission, such as double channel training based method [6] and two-way training based method [7]; 3) \emph{statistic feature (SF) based method} [8], [9] is an effective approach to detect the attack by extracting effective statistic features, such as the subspace dimension [8] and energy characteristic [9]. However, existing statistic features are not robust to discriminate between spoofing attack and other causes of fluctuations induced by the legitimate communication itself. Moreover, the SF-based method relies on the prior knowledge about signal features based on the assumption that all channel coefficients are independent and identical distributed (i.i.d.) Gaussian random variables, which is not applicable to the fast fading in link and system-level simulations for 5G NR in 3GPP standards [10].

Motivated by the aformetioned issues, we propose a spatial sparsity structure based method to detect the DMRS spoofing inspired by the idea of sequential change point detection (SCPD) [11], which can tackle the above three major issues as follows: 1) The proposed sequential sparsity structure anomaly detection (SSSAD) method is more practical and easier to implement, since no modification of communication protocol and additional auxiliary transmission are required; 2) The extracted spatial sparsity structure of the channel can distinguish the normal case and the DMRS spoofing with low probability of false alarm. Because the facing scatterers are different, the legitimate user and the attacker with different locations will result in different angle of arrivals (AoAs) at the BS, and finally appear as the change of the spatial sparsity structure of the channel due to the contaminated channel estimation caused by the DMRS spoofing; 3) The proposed SSSAD method can work in a real time manner to quickly detect abnormalities and does not rely on the prior knowledge of the attacker, because it only cares the abrupt change in the spatial sparsity structure of the channel. In addition, we adopt the clustered delay line (CDL) based channel model from 3GPP standards [10], which is more suitable for system level simulations of 5G NR. Simulation experiments confirm the effectiveness of our detection method.

\section{System Model}

We consider an uplink time division duplex (TDD) communication system, where an $M$-antenna BS serves $K$ single-antenna legitimate users using DMRS for channel estimation as part of coherent demodulation of PUSCH. Since the initial access and DMRS transmission protocol are publicly known, as the worst case, we consider that DMRS assignments to users are known by the attacker. As illustrated in Fig. 1, an intelligent attacker can attack the PUSCH by sending the same DMRS of a victim legitimate user, i.e., DMRS spoofing. In this correspondence, we only consider the low mobility scenarios with the CDL channel model. The cases with high mobility scenarios will correspond to a totally different type of DMRS design for the 3GPP 5G NR [12], [13] and left for future research. In addition, we mainly consider the situation where the attacker is targeted at a single specific user as in [7]-[9]. Note that our method based on the channel estimation result of each user also applies to multiusers DMRS spoofing (see our previous work [14] for more details).

According to the NR specification, each user is assigned a preconfigured DMRS, which is employed for coherent demodulation of the PUSCH is generated using multiple-root Zadoff-Chu (ZC) sequences, see Section 5.2.2 of [12] for more details. The ZC sequences exhibit an excellent cross-correlation property that enables low-complexity multipath channel estimation. Denoting an ${N_{RS}}$ length ZC sequence with root $r = \left\{ {1,\cdots ,{N_{RS}} - 1} \right\}$ as ${\bm z_r}$, its $j$th element is given as ${z_r}\left( j \right) = \frac{1}{{\sqrt {{N_{RS}}} }}{e^{\left( {{{ - i\pi rj\left( {j + 1} \right)} \mathord{\left/
 {\vphantom {{ - i\pi rj\left( {j + 1} \right)} {{N_{RS}}}}} \right.
 \kern-\nulldelimiterspace} {{N_{RS}}}}} \right)}},j = 1,\cdots ,{N_{RS}}$. Based on ${\bm z_r}$, multiple reference sequences can be generated from a single-root ZC sequence by cyclically shifting it by a shift size greater than delay spread length. Denote the set of reference sequences as $ \bm {\Phi}  \triangleq \left\{ {{\bm p_1},\cdots ,{\bm p_Q}} \right\}$, i.e., a pool of $Q$ preambles, the $Q$ preambles can be generated from ZC sequences, then the BS allocates these sequences to the legitimate users. Let ${\bm p_k} \in {{\mathbb C}^{{N_{RS}} \times 1}}$ be the transmitted reference sequence by the $k$th user ($k = 1, \cdots ,K$). Without loss of generality, we assume the $k$th user is the target of attack.

 \begin{figure}[t]
 	\centering
 	\includegraphics[width=3 in]{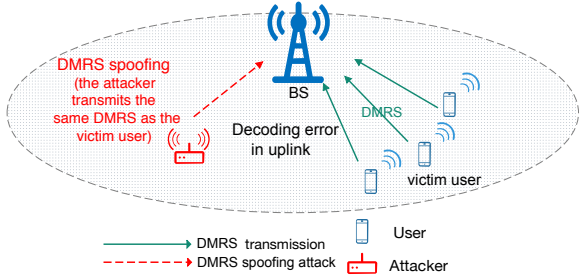}
 	\caption{ DMRS spoofing.}
 	\label{w1}
 \end{figure}
 
We use ${\bm h_{k,m}} \in {{\mathbb C}^{\tau  \times 1}}$ to denote the time-domain channel from the $k$th user to $m$th BS antenna, for $m = 1,2, \cdots ,M$. In particular, ${\bm h_{k,m}} \triangleq {\left[ {{h_{k,m,1}},{h_{k,m,2}}, \cdots ,{h_{k,m,\tau }}} \right]^T}$, where ${h_{k,m,t}}$ represents a multipath channel component for the $t$th tap with delay-spread length $\tau$, for $t = 1,2, \cdots ,\tau $. Similarly, we use ${\bm g_{A,m}} \triangleq {\left[ {{g_{A,m,1}},{g_{A,m,2}}, \cdots ,{g_{A,m,\tau }}} \right]^T} \in {{\mathbb C}^{\tau  \times 1}}$ to denote the channel between the attacker and $m$th BS antenna. We further have that ${\bm h_k} \triangleq \left[ {{\bm h_{k,1}},{\bm h_{k,2}}, \cdots ,{\bm h_{k,M}}} \right]$ and ${\bm g_A} \triangleq \left[ {{\bm g_A}_{,1},{\bm g_A}_{,2}, \cdots ,{\bm g_A}_{,M}} \right]$. Note that we consider a CDL-D based channel model in order to validate the effectiveness of the proposed detection method in system-level simulations for 5G NR. A detailed listing of various related information of the delay and angular spreads used is provided in Technical Report (TR) 38.901 (TABLE 7.7.1-4) [10].

\section{DMRS Spoofing Detection via SSSAD Method}

In this section, we first describe the DMRS spoofing model, then introduce detection principle of our proposed SSSAD method, and details are given for several processing stages involved in the core sequential anomaly detection by using the spatial sparsity structure of the channel.

\subsection{DMRS Spoofing}

In DMRS spoofing, as shown in Fig. 2, the attacker sends the same DMRS as the victim user to disturb the uplink channel estimation. DMRS needs to be transmitted in specific time-frequency resources blocks (RBs) scheduled for PUSCH. Once DMRS suffering such attack, the block rate of the whole cell will be degraded, since the coherent demodulation of PUSCH of all paired users are also likely to be affected especially when the bandwidth occupation is sparse. Then, it will result in severe denial of services. Moreover, the DMRS spoofing will seriously affect PUSCH coverage due to the channel estimation errors, thereby limiting the uplink coverage of the system and affects the performance of cell-edge users.

 \begin{figure}[t]
 	\centering
 	\includegraphics[width=2.5 in]{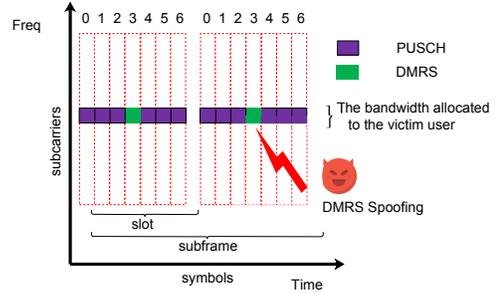}
 	\caption{5G NR frame structure of the victim user under DMRS spoofing.}
 	\label{w1}
 \end{figure}

Considering the basic orthogonal frequency-division multiplexing (OFDM) procedure in NR, the received time-domain signal vector at the $m$th receive antenna in the BS under the DMRS spoofing, denoted by $\bm y_m^{TD}$, is given by
\begin{align}
\bm y_m^{TD} = \sum\limits_{k = 1}^K {\sqrt {{P_k}} {\bm p_k} \otimes {\bm h_{k,m}} + \sqrt {{P_A}} {\bm p_k} \otimes {\bm g_{A,m}} + } \bm w_m^{TD},
\end{align}
where $P_k$ is the DMRS transmission power of the $k$th user, $P_A$ is the attacking power, $\otimes$ denotes a circular convolution operation, and $\bm w_m^{TD}$ is the additive white Gaussian noise (AWGN) at the BS with each element being distributed as $\mathbb{C}\mathbb{N}\left( {0},\sigma^2 \right)$. By applying fast Fourier transform (FFT), the received frequency-domain signals at the $m$th receive antenna in the BS can be written as
\begin{align}
&\bm y_m^{FD}\left( l \right) = \sum\limits_{k = 1}^K {{\rm diag}({\sqrt {N_s{P_k}} \bm F{{\left[ {\bm h_{k,m}^T{\text{ }}{\bm 0_{1 \times \left( {N_s - \tau } \right)}}} \right]}^T}} ){\bm p_k}\left( l \right)+} \nonumber \\
&{\rm diag} ( {\sqrt {N_s{P_A}} \bm F{{\left[ {\bm g_{A,m}^T{\text{ }}{ \bm 0_{1 \times \left( {N_s - \tau } \right)}}} \right]}^T}} ){\bm p_k}\left( l \right) + \bm w_m^{FD}\left( l \right),
\end{align}
where $\bm F$ denotes unitary discrete Fourier transformation (DFT) matrix, $l$ is the number of samples ($l = 1,2, \cdots ,L$), and $N_s$ is the number of subcarriers. $\bm w_m^{FD}$ is the DFT projection of the random vector $\bm w_m^{TD}$. Then, by combining the received signals across $M$ antennas, the signal received at the BS can be reformulated as
\begin{align}
\bar {\bm y}\left( l \right) = \sum\limits_{k = 1}^K {\sqrt {{P_k}} {\bm p_k}\left( l \right)\bar {\bm h}_k^{}}  + \sqrt {{P_A}} {\bm p_k}\left( l \right)\bar {\bm g}_A^{} + \bar {\bm w}\left( l \right),
\end{align}
where $\bar {\bm y} \left( l \right) \triangleq \left[ {{{\left( {\bm y_1^{FD}\left( l \right)} \right)}^T}, \cdots ,{{\left( {\bm y_M^{FD}\left( l \right)} \right)}^T}} \right]$, $\bar {\bm w}\left( l \right) \triangleq \left[ {{{\left( {\bm  w_1^{FD}\left( l \right)} \right)}^T}, \cdots ,{{\left( {\bm w_M^{FD}\left( l \right)} \right)}^T}} \right]$ is the additive noise at the BS. $\bar {\bm h}_k^{} \triangleq\left[ {{{\left( {\bar {\bm F}\bm h_{k,1}^{}} \right)}^T} \cdots ,{{\left( {\bar {\bm F} \bm h_{k,M}^{}} \right)}^T}} \right]$, $\bar {\bm g}_A^{} \triangleq \left[ {{{\left( {\bar {\bm F} \bm g_{A,1}^{}} \right)}^T}, \cdots ,{{\left( {\bar {\bm F} \bm g_{A,M}^{}} \right)}^T}} \right]$, $\bar {\bm F} \triangleq   {\bm F}\left( {:,1:\tau } \right)  \sqrt N$. Without awaring the DMRS spoofing attack, the BS will make a mistake on the estimation of $\bar {\bm h}_k^{}$ according to the least square (LS) principle as 
\begin{align}
\hat {\bm h}_k(l) = \bar {\bm h}_k + \rho \bar {\bm g}_A^{} + \bm w\left( l \right),
\end{align}
where $\rho  \triangleq \sqrt {{P_A}}/{{P_k}} $, and $\bm w\left( l \right) \triangleq {{\bar {\bm w}\left( l \right){{\left( {{\bm p_k}\left( l \right)} \right)}^ * }} / {\left( {{N_{RS}}\sqrt {{P_k}} } \right)}}$ is the equivalent noise vector with distribution $\mathbb{C}\mathbb{N}\left( \bm{0},(\sigma^2 / ({{N_{RS}}\sqrt {{P_k}} })){\bm{I}_{NM}} \right)$. Consequently, the channel estimation errors will severely degrade the decoding performance.

\subsection{Detection Framework of SSSAD}

Define the already collected at the channel estimation in the $t$th subframe as ${s}^{[t]}(l)\triangleq (\hat {\bm h}^{}(l))^H\hat {\bm h}^{}(l)$, we drop the superscript of index $k$ for brevity. Our goal is to detect whether the newly arriving samples ${s}^{[t+1]}(l)$ in the $(t+1)$th subframe are anomalous or not in a reliable manner, that is to say, we plan to solve the following detection problem by utilizing the spatial sparsity structure of the channel from the monitoring samples ${s}^{[t]}(l)$

\begin{align}
{s}^{[t+1]}(l)=\left\{
\begin{aligned}
&{ (\bar {\bm h})^H\bar {\bm h}+\sigma^2 },&& \rm {no\ {attack}},\\
&{{[ \bar {\bm h} ; \bar {\bm g}_A ]^H} 
\begin{matrix}
   \left[ {\matrix
   { 1 } & {  \rho }  \\ 
   {  \rho } & {\rho^2 }  \\ 
 \endmatrix } \right]
\end{matrix}
[ \bar {\bm h}; \bar {\bm g}_A ]+\sigma^2},&& \rm {DMRS\  spoofing}.
\end{aligned}\right. \nonumber
\end{align}

Next, we present our SSSAD method based on the above detection problem. In practice, the BS equipped with hundreds of antennas is erected high enough to ensure the spatial resolution and reduce the angular spread of the incident signals, the \emph{spatial sparsity of the channel} is easy to realize. Each element in ${\bm h_{k,m}}$ or ${\bm g_{A,m}}$ indicates the distribution of gain in a specific direction, the spatial sparsity of the channel specifically refers to only a small number of elements that are significant, i.e., that point to the AOA of the signal.

Motivated by this available physical attribute, in SSSAD, we extract the spatial sparsity of the channel from the monitoring samples and the newly arriving samples. If the DMRS spoofing exists, the spatial sparsity structure of the channel found by our SSSAD method will be remarkably affected, since the legitimate user and the attacker with different locations will result in different AOAs, and the DMRS spoofing will lead to a combination structure of spatial sparsity from the legitimate channel and the illegitimate channel as described in (4). Specifically, we define the indicator of the spatial sparsity of the channel in the $t$th subframe as $\bm{\phi }^{[t]}$. Given $L$ sample observations in the $t$th subframe ${s}^{[t]}(l)$, and the sparsity extraction for the DMRS spoofing detection can be expressed as the following sparse feature retrieval problem
\begin{align} 
\mathop {\mathrm {min}} &~~~\frac{1}{L}\sum _{l=1}^{L}\bigl [{s}^{[t]}(l)- (\zeta _l({\bm{\phi }^{[t]}}))^H \zeta _l({\bm{\phi }^{[t]}})- \xi _L({\bm{\phi }^{[t]}})\bigr ]^2, \\
s.t. &~~~{\Vert {\bm{\phi }^{[t]}}\Vert _0\le r},
\end{align}
where the objective function in (5), denoted by $\mathcal {L}({\bm{\phi }^{[t]}})$, is the sample version of the variance loss function ${\mathbb V}{\text{ar}}[{s}^{[t]}(l)- (\zeta _l({\bm{\phi }^{[t]}}))^H \zeta _l({\bm{\phi }^{[t]}})]$, $r$ is sparse constraint of $\bm{\phi }^{[t]}$, $\zeta _l({\bm{\phi }^{[t]}})\triangleq {\bm{h}^{H}_I}{(l)} {\bm{\phi }^{[t]}}$ with ${\bm{h}_I}$ generated from standard scaled CDL-D model and $\xi _L({\bm{\phi }^{[t]}})\triangleq \mu _L - \Vert {\bm{\phi }^{[t]}}\Vert ^2$ with $\mu _L= L^{-1}\sum _{l=1}^L{{s}^{[t]}(l) }$. 

We emphasize that we use the spatial sparsity structure of the channel to implement more robust anomaly detection and adopt a real time sequential manner without assuming any attacker’s information. The proposed SSSAD method can be efficiently carried out by the following \emph{Stage 1: sparsity feature extraction} and \emph{Stage 2: sequential anomaly detection}.

Stage 1: sparsity feature extraction. In order to solve the minimization problem of $\mathcal {L}({\bm{\phi }^{[t]}})$ efficiently, we propose a gradient-based algorithm. \emph{1) Update of $\bm{\phi }^{[t]}$:} Considering the cardinality constraint of $r$ is usually unknown, we utilize an adaptive threshold function to the estimate after a gradient descent step, which is possible to determine the sparsity of the estimate in an iterative manner. Similar to [15], for $n=1,2,\cdots, N$, the threshold value at the $n$th iteration can be computed by
\begin{align} 
\delta ({\bm{\phi }_{(n)}^{[t]}}) = & \alpha \biggl \{ \kappa \sum _{i = 1}^L [{s}^{[t]}(l) - (\zeta _l({\bm{\phi }_{(n)}^{[t]}}))^H \zeta _l({\bm{\phi }_{(n)}^{[t]}})- \nonumber \\ & \mu _L + \Vert {\bm{\phi }_{(n)}^{[t]}} \Vert ^2] ^2(\zeta _l({\bm{\phi }_{(n)}^{[t]}}))^H \zeta _l({\bm{\phi }_{(n)}^{[t]}})\biggr \}^{1/2}, 
\end{align}
where $\alpha$ is the tuning parameter and $\kappa \triangleq \frac{\log (M \tau L)}{(M\tau)^2}$. Then, we perform a thresholded gradient descent step
\begin{align} 
{\bm{\phi }_{(n+1)}^{[t]}}={\mathcal {D}}_{{\beta } \cdot \delta ({\bm{\phi }_{(n)}^{[t]}})}\bigl [ {\bm{\phi }_{(n)}^{[t]}} - {\beta } \nabla \mathcal {L} _l({\bm{\phi }_{(n)}^{[t]}})\bigr ], 
\end{align}
where $\beta >0$ is the step size, $[{\mathcal {D}}_{\delta }({\bm{z}})]_j = {\bm{z}}_j\cdot \bm {1}(|{\bm{z}}_j| \ge \delta )$ is the operator for hard-thresholding, and $\nabla \mathcal {L} _l({\bm{\phi }_{(n)}^{[t]}})=\sum ^L_{l=1}\bigl [{s}^{[t]}(l) - \zeta _l({\bm{\phi }_{(n)}^{[t]}})- \xi _L({\bm{\phi }_{(n)}^{[t]}})\bigr ] \bigl ( \bm {I}_{M\tau}- {\bm{h}^{}_I}{(l)}{\bm{h}^{H}_I}{(l)}\bigr ){\bm{\phi }_{(n)}^{[t]}}$.

After updating ${\bm{\phi }_{(n)}^{[t]}}$ for $N$ iterations for some sufficiently large $N$, we take the last estimate ${\bm{\phi }_{(N)}^{[t]}}$ as the final estimate of $ {\hat {\bm{\phi }}}^{[t]}$. To run the gradient step in (8), we need to obtain a good initializer, which can be easily done by thresholded spectral method [15]. Applying spectral methods to the empirical equivalent of matrix $\mathbb {E}[ ({s}^{[t]} - \mu_L ) {\bm{h}^{}_I}{}{\bm{h}^{H}_I}{} ]$ based on samples is a normal process. Due to the fact that $\mathbb {E}[ ({s}^{[t]} - \mu_L ) {\bm{h}^{}_I}{}{\bm{h}^{H}_I}{} ]$ is a high-dimensional matrix, the largest eigenvector in this sample matrix may have a considerable estimation error. To tackle this problem, we use spectral methods on a submatrix through the following step. We first perform a thresholding process to choose a subset of coordinates ${\hat{\Gamma}}_0$ by
\begin{align} 
{\hat{\Gamma}}_0 = \biggl \{ i\in [M\tau]:\biggl | \frac{1}{L}\sum _{l=1}^L{s}^{[t]}(l)\bigl [{\bm{h}^{i}_I}{(l)}-1\bigr ]\biggr |>\gamma  \biggr \}, 
\end{align}
where $\gamma\triangleq \sqrt{ \log (M \tau L)/ (M \tau) }$. Since the diagonal entries of the matrix $\mathbb {E}[ {s}^{[t]} ({\bm{h}^{}_I}{}{\bm{h}^{H}_I}{}-  \bm {I}_{M\tau})]$ on the support of $ {\hat {\bm{\phi }}}^{[t]}$ are non-zero, which motivates the thresholding step in (7). $L^{-1} \sum _{l=1}^L{s}^{[t]}(l) [{\bm{h}^{i}_I}{(l)}-1 ]$ is close to its expectation for all $i \in [M\tau]$ when $L$ is sufficiently large. All above descriptions mean that the thresholding process creates a ${\hat{\Gamma}}_0$ that is a subset of the true support $\mathrm {supp}({\hat {\bm{\phi }}}^{[t]})$, which is similar to the diagonal thresholding approach for sparse principle component analysis.

\emph{2) Initialization:} We use $\mu_L$ to denote the sample average of ${s}^{[t]}_{(1)}, \ldots , {s}^{[t]}_{(L)}$ and define ${\bm{Z}}\triangleq \frac{1}{L}\sum _{l= 1}^{ L} ({s}^{[t]}(l)-\mu _L){\bm{z}^{}}{(l)}{\bm{z}^{H}}{(l)}$. Note that the $|{\hat{\Gamma}}_0| \times |{\hat{\Gamma}}_0|$ non-zero entries of ${\bm{Z}}$ form the submatrix of $L^{-1} \sum _{l=1}^{L} ({s}^{[t]}(l) - \mu _L){\bm{h}^{}_I}{(l)}{\bm{h}^{H}_I}{(l)}$ with both rows and columns in ${\hat{\Gamma}}_0$. Let $\bm{v}$ be the eigenvector of ${\bm{Z}}$ that corresponds to the eigenvalue with the biggest magnitude. Finally, the initial estimator ${\bm{\phi }_{(0)}^{[t]}}$ is given by ${\bm{\phi }_{(0)}^{[t]}}= \bm{v}\sqrt{| \psi _L| / 2}$, $\psi _L = {\frac{1}{L}\sum _{l=1}^{ L} {s}^{[t]}(l) ( {\bm{h}^{H}_I}{(l)} {\bm{v }}  )^H( {\bm{h}^{H}_I}{(l)} {\bm{v }}  ) - \mu _L}$, and the iterations using thresholded gradient descent, as defined in (8), outputs the estimate of $ {\hat {\bm{\phi }}}^{[t]}$. Since they are indistinguishable from the error, the coordinates outside the effective support of ${\hat {\bm{\phi }}}^{[t]}$ can be safely omitted in the initialization. It is critical to complete the thresholding step in (9) before building $\bm{Z}$. The noise introduced to the spectral estimator after the thresholding step is roughly proportional to $|{\hat{\Gamma}}_0|$ linearly due to $|{\hat{\Gamma}}_0|\le r$, which is substantially less than the dimension of sample observations. Finally, we can get a proper initial estimator. 

Stage 2: sequential anomaly detection. After extraction of the spatial sparsity of the channel, we can perform sequential anomaly detection in a real time manner to quickly detect the DMRS spoofing. \emph{1) Sequential process:} At each time when a new sample observation is obtained, the detector makes a decision based on all the sample observations collected at hand. Combined with the DMRS spoofing detection in this work, if no change in the spatial sparsity structure of the channel is detected, then the detector moves to the next subframe instant, collecting new samples and making a new decision. We can see that with sequential process, our detection method works in a real time manner and does not rely on the prior knowledge of the attacker. \emph{2) Anomaly detection:} To exploit the change of the spatial sparsity structure of the channel between the monitoring samples ${s}^{[t]}(l)$ in the $t$th subframe and the newly arriving samples ${s}^{[t+1]}(l)$ in the $(t+1)$th subframe, we define a metric $c\triangleq |\frac{ \hat {\bm{\phi }}^{[t]}\hat {\bm{\phi }}^{[t+1]}} {\Vert {\hat {\bm{\phi }}^{[t]}}\Vert \Vert \hat {\bm{\phi }}^{[t+1]} \Vert}|$ to capture the sparsity structure change. If the newly arriving samples ${s}^{[t+1]}(l)$ result in a large change of the spatial sparsity structure of the channel and thus $c$ is lower than the pre-defined detection threshold $\eta$, we will claim that the DMRS spoofing exists. This is because if the DMRS is only coming from the legitimate user, the current obtained sparsity structure of the channel and the recorded one should be similar since they have the same physical location and AOA. Conversely, the similarity would be degraded under the spoofing attack, in which there are more transmitters in the channel, i.e., the contaminated channel estimation caused by the DMRS spoofing. The detection threshold $\eta$ can be set empirically through simulations. Based on the probability theory and statistics, we use the cumulative distribution function (CDF) to help to set up $\eta$. As nearly all the $c$ is larger than 0.92, thus we set $\eta=0.92$ in all simulations. 

As a result, we can use the similarity between the recorded sparsity structure of the channel and the current one to detect the DMRS spoofing by the combination of Stages 1 and 2.

\section{Numerical Results}

In this section, we present simulation results to show the detection performance of the proposed SSSAD method. Based on the typical 5G NR OFDM transmission scheme proposed by 3GPP [16], we adopt CDL-D model according to TR 38.901 (TABLE 7.7.1-4) [10]. The parameters for setting the communication scenario are the ground city macro-cell parameters [10]. The user and the attacker are uniformly distributed in a circular region with the BS located at the center and the inner and outer radiuses of 100 m and 120 m, respectively. For the DMRS structure, carrier bandwidth of user may choose from 16RB or 4 RB. We set $M=64$, $K=16$, $\alpha=15$, and assume $P_{U} = P_{k} = 10$dBm, $k = 1,2,\cdots,K$. We set the received jammer-to-signal ratio (JSR) $P_{A}/P_{U}=0$dB. We use the received signal-to-noise ratio (SNR) $P_{U}/{\sigma^2}$ to evaluate the impact of the noise power level on the detection performance. The subspace dimension (SD) based method [8] and the energy detector (ED) based method [9] are considered for comparison. 

\begin{figure}[t]
\begin{center}
\subfigure[$P_{D}$ v.s. $P_{FR}$, SNR$=5$dB.]{
    \label{dJA:sub1} 
    \includegraphics[width=1.9 in]{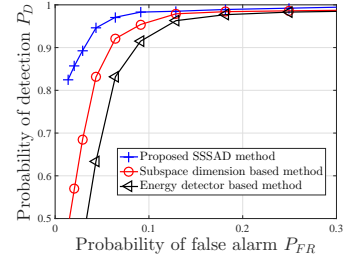}}
    \hspace{.01in}
\subfigure[$P_{D}$ v.s. $P_{FR}$, SNR$=-5$dB.]{
    \label{dJA:sub2} 
    \includegraphics[width=1.9 in]{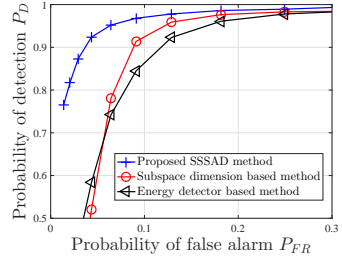}}
    \vspace{.01in}
\caption{\small The comparison of ROC of different methods, where carrier bandwidth of user is 16 RB.}\label{Fig:dJA}
\end{center}
\vspace{-5mm}
\end{figure}

To evaluate the detection performance of different methods, the receiver operating characteristic (ROC) curves are plotted under different parameters. In Fig. 3, we plot the detection probability $P_{D}$ versus false alarm probability $P_{FR}$ of different methods under 16RB, i.e., these detectors collect together all the observations in 16RB and then make the final decisions. In the relatively high SNR region, Fig. 3(a) reveals that the detection accuracy of the proposed SSSAD method is better than the ED-based method and SD-based method for a given $P_{FR}$. The SD-based method can achieve better performance than the ED-based method, but demonstrates a poor performance with the increasing requirement of $P_{FR}$. When the SNR is low, as shown in Fig. 3(b), we observe that both the ED-based and SD-based methods exhibit poor detection performance. Especially, the detection performance of the SD-based method declines sharply, which greatly limits its scope of application. By contrast, the detection probability of our SSSAD method only has a small decline and appears to be robust enough to different noise power levels. Moreover, the detection performances of both the ED-based and SD-based methods decline noticeably under CDL-D channel model, further evaluations in practical application are required. If the number of observations is small, in Fig. 4, we plot the ROC curves under 4RB. Compared with Fig. 3, when the number of observations available for detection decreases, the decline in the detection accuracy is particularly apparent for the ED-based method and SD-based method. Particularly for the ED-based method, as shown in Figs. 4(a) and 4(b), its detection probabilities drop rapidly and show great fluctuation for both high and low SNR. So, the ED-based method needs the support of large observations, and not applicable to detect DMRS spoofing for per transmission time interval in practice.

In summary, Figs. 3 and 4 reveal that the proposed SSSAD method can detect the occurrence of DMRS spoofing more accurately under different number of observations and SNR, and more suitable for CDL model for 5G NR in 3GPP standards than the benchmarks ED-based and SD-based methods.

\begin{figure}[t]
\begin{center}
 \subfigure[$P_{D}$ v.s. $P_{FR}$, SNR$=5$dB.]{
    \label{dJA:sub1} 
    \includegraphics[width=1.9 in]{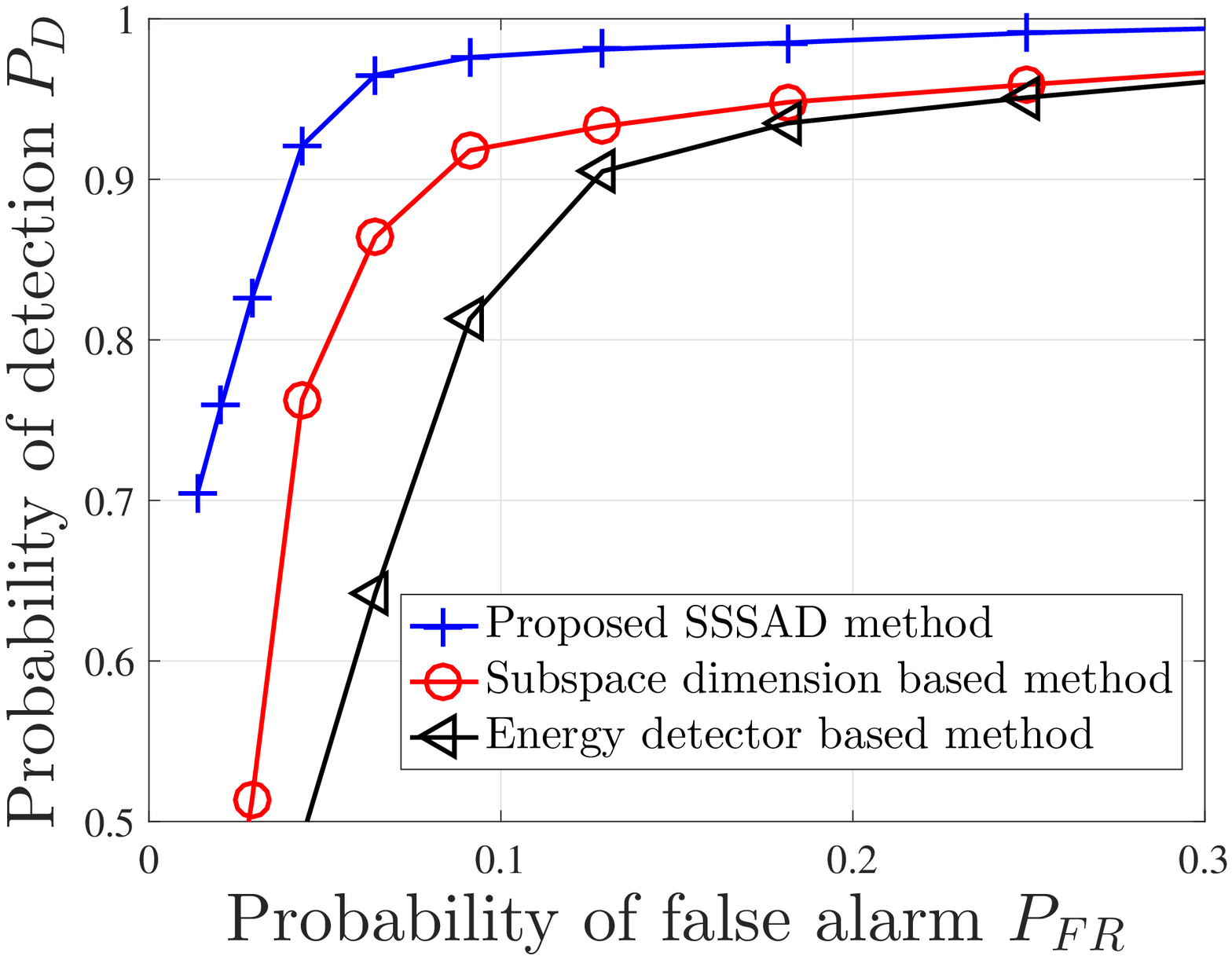}}
    \hspace{.01in}
 \subfigure[$P_{D}$ v.s. $P_{FR}$, SNR$=-5$dB.]{
    \label{dJA:sub2} 
    \includegraphics[width=1.9 in]{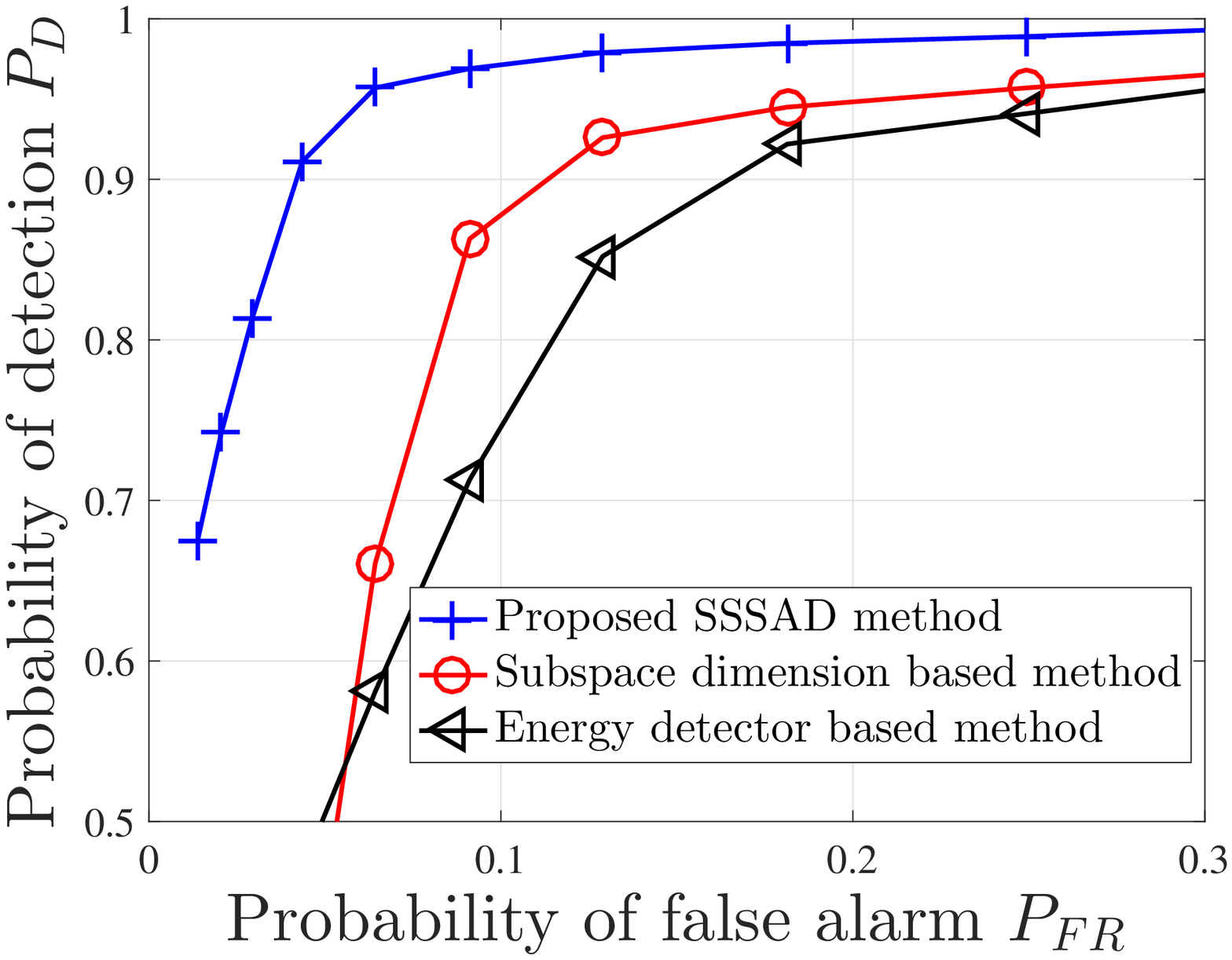}}
    \vspace{.01in}
 \caption{\small The comparison of ROC of different methods, where carrier bandwidth of user is 4 RB.}\label{Fig:dJA}
\end{center}
\vspace{-5mm}
\end{figure}

\section{Conclusion}
\label{Sec:Conclusion}
In this correspondence, we have considered the DMRS spoofing, and an efficient SSSAD method was developed to detect the DMRS spoofing under the CDL-based channel model from 3GPP standards. Our key insight is that an abrupt change in the spatial sparsity structure of the channel will occur once the DMRS spoofing exists. To evaluate the detection performance, ROC curves under different parameters were plotted. Simulations have shown that the performance of our method outperforms the ED-based and SD-based methods.

\end{document}